\documentclass[]{aastex}
\usepackage{emulateapj5,onecolfloat,graphicx}
  \usepackage[hyperindex,breaklinks]{hyperref}
\def\GCS{{\small GCS}}
\def\Hipp{{\it Hipparcos}}
\def\RAVE{{\small RAVE}}
\def\SDSS{{\small SDSS}}

\def\eg{{\it e.g.}}
\def\etal{{\it et al.}}

\def\ie{{\it i.e.}}
\def\vs{{\it vs.}}
\def\Teff{{$T_{\rm eff}$}}
\def\FeH{[Fe/H]}
\long\def\Ignore#1{\relax}

\slugcomment{Accepted to appear in ApJ May 29, 2012, first submitted Sep 30, 2011}

\shorttitle{Radial mixing}
\shortauthors{Yu \etal}

\begin{document}

\twocolumn[
\title{A Test for Radial Mixing Using Local Star Samples}

\author{Jincheng Yu}
\affil{Key Laboratory for Research in Galaxies and Cosmology, Shanghai Astronomical Observatory, \\
   Chinese Academy of Sciences, 80 Nandan Road, Shanghai 200030, China \\
Graduate University of the Chinese Academy of
  Sciences, 19A Yuquanlu, Beijing 100049, China}
\email{yujc@shao.ac.cn}

\author{J. A. Sellwood and Carlton Pryor}
\affil{Department of Physics and Astronomy, Rutgers University, \\
    136 Frelinghuysen Road, Piscataway, NJ 08854}
\email{sellwood,pryor@physics.rutgers.edu}

\author{Li Chen and Jinliang Hou}
\affil{Key Laboratory for Research in Galaxies and Cosmology, Shanghai Astronomical Observatory, 
   Chinese Academy of Sciences, 80 Nandan Road, Shanghai 200030, China}
\email{chenli,houjl@shao.ac.cn}


\begin{abstract}
We use samples of local main-sequence stars to show that the radial
gradient of \FeH\ in the thin disk of the Milky Way decreases with
mean effective stellar temperature.  Many of these stars are visiting
the solar neighborhood from the inner and outer Galaxy.  We use the
angular momentum of each star about the Galactic center to determine
the guiding center radius and to eliminate the effects of epicyclic
motion, which would otherwise blur the estimated gradients.  We
interpret the effective temperature as a proxy for mean age, and
conclude that the decreasing gradient is consistent with the
predictions of radial mixing due to transient spiral patterns.  We
find some evidence that the trend of decreasing gradient with
increasing mean age breaks to a constant gradient for samples of stars
whose main-sequence life-times exceed the likely age of the thin disk.
\end{abstract} 

]

\section{Introduction}
A radial abundance gradient in the Milky Way disk today is
well-established from analyses of the interstellar medium
\citep{Shav83,Bals11}.  Metal abundances are believed to increase over
time and most chemical evolution models
\citep[\eg,][]{Bois99,Hou00,Chia01,Naab06,Fu09} predict that a radial
abundance gradient is created early and may even have been steeper in
the past.  This picture is consistent with ``inside-out'' disk
formation, in which the stellar population at increasing radii is
younger and more metal poor \citep[\eg][]{deJo96,Muno07}.

The chemical composition of a star is that of the cloud of gas from
which it formed.  Abundance gradients are therefore also observed in
the disk of the Milky Way in stars
\citep{Nord04,Hayw08,Luck11,Schl11}, in planetary nebulae
\citep{Maci07,SH10}, and in open star clusters \citep{Frie02,Chen03}.
While all agree that abundances in the thin disk decrease outwards,
the precise slope differs from sample to sample and element measured.

The orbits of disk stars generally become more eccentric over time
\citep{Wiel77,Holm09,AB09}, which ``blurs'' measurements of the
metallicity gradient when the instantaneous radii of older stars are
used.  Eccentric motion of any amplitude, projected to the mid-plane
of an axisymmetric potential, can be described as a retrograde
epicycle about a guiding center that orbits at a constant angular rate
\citep{BT08}.  The guiding center, or home, radius of a disk star is
directly related to the $z$-component of the angular momentum, $L_z$,
about the Galactic center and, since angular momentum is conserved,
can be determined from any point around the epicycle.  Thus
measurement of the angular momenta of stars allows us to eliminate the
blurring effects of epicyclic motions and simplifies the
interpretation of metallicity gradients.  \citet{Nord04} used the mean
Galacto-centric radius for the same reason, which requires integrating
the orbit in some adopted potential -- angular momentum is more
direct.

In the absence of radial mixing, chemical evolution models in which
the metallicity of the ISM increases at each radius over time would
predict a tight correlation between the metallicity and age for stars
of a given home radius.  This prediction is not supported by the data,
and the metallicity distribution of all but the youngest stars shows a
broad distribution \citep{Edva93,Nord04}, which persists even after
correcting for epicyclic blurring \citep{Hayw08}.

\citet{SB02} argued that stars in a disk galaxy are shuffled in radius
over time by the effects of transient spiral arms.  They showed that a
star near the corotation resonance of a spiral could gain or lose
enough angular momentum for its home radius to change by up to 2~kpc,
leaving the star at its new mean galacto-centric distance with no
increase in its epicyclic amplitude.  The resulting radial mixing is
described as ``churning'' and their result has been confirmed and
extended \citep{Rosk08a,Rosk08b,Loeb11,Minc11,Bird11,SSS12}.  The
broad spread of metallicities among stars having common ages and home
radii today is then naturally explained as reflecting the past
gradient of metallicity of the ISM at their various birth radii.
\citet{SB09} provide a model for chemical evolution of the Milky Way
that was the first to include radial mixing.  \citet{Loeb11} find good
agreement between the predictions of their simulations, that manifest
substantial radial mixing, with the metallicity distribution of stars
from the \SDSS\ \citep{Ivez08}; \citet{Lee11} find other evidence
supporting radial migration.

An additional consequence of radial mixing, or churning, is that the
radial metallicity gradient of a generation of stars is gradually
flattened over time.  \citet{Nord04} and \citet{Hayw08}, from local
stars suggest that the gradient may become shallower with increasing
age.  \citet{Casa11} report little evolution of the radial metallicity
gradient over the past $\sim 5~$Gyr, using infrared magnitudes and
improved models to revise the age estimates for the stars in the
\citet{Nord04} sample.  Estimating the age of an individual star
\citep[see][for a recent review]{Sode10} is a delicate art and the
results even for well-observed main-sequence stars can be contentious
\citep[\eg][]{Reid07,Holm07}.  Furthermore, the evidence from star
clusters \citep{Frie02,Chen03} suggests that the radial metallicity
gradient may steepen with age, in line with inside-out models of
galaxy formation.  \citet{Maci07}, from a study of planetary nebulae
abundance data, also find evidence for a steeper gradient among their
older objects, although \citet{SH10} reach the opposite conclusion
that older PNe show a shallower gradient.  We comment on these
apparently conflicting results in \S4.2.

Here we study the age-dependence of the metallicity gradient using a
sample of stars in the solar neighborhood.  Instead of estimating ages
of individual stars, we adopt the effective temperature of a
collection of main sequence stars as a proxy for their mean age, in
the same manner that \citet{AB09} used color.  Since the main-sequence
lifetime of a star is shorter for hotter stars, the mean ages of main
sequence disk stars grouped by effective temperature must be lower for
the hotter groups.  (Of course, this argument requires the reasonable
assumptions that stars have been forming with an approximately
constant IMF and at a roughly uniform rate over the lifetime of the
Milky Way disk.)  The trend of increasing mean age with decreasing
temperature must change to a constant at the point at which the
main-sequence lifetimes of stars exceed the age of the disk.  Note
that the main-sequence turn-off for stars of age 10~Gyr occurs for
$5\,500 \la T_{\rm eff} \la 6\,000\;$K over the range $0.4
\ga\;$\FeH$\;\ga-0.5$ \citep{Dema04}.

As explained above, we eliminate epicyclic blurring by estimating the
specific angular momentum of each star, which requires knowledge of
its full 6D phase space coordinates: \ie\ distance and proper motion,
as well as radial velocity and sky position.  A radial velocity can be
measured spectroscopically at any distance, but a reliable distance
can be obtained only for nearby stars.  Furthermore, distance
uncertainties factor into velocity components in the sky plane adding
to the desirability of restricting attention to stars close to the
Sun.  Also, by focusing on nearby stars, any variations in $L_z$
across our sampling volume due to possible departures from axisymmetry
in the Galactic potential will be negligible.

A benefit of stellar epicycle excursions is that they bring stars to
the solar neighborhood.  Thus a sample of very local stars will span a
significant range of specific angular momenta and, therefore, home
radii.

In this paper, we assemble a sample of local, main-sequence stars
having estimated effective temperatures, metallicities and full space
motions.  We draw these stars from three sources as described in the
next section.

\section{Star samples}
\subsection{Geneva-Copenhagen sample}
\label{GCSsample}
The Geneva-Copenhagen survey of nearby stars \citep{Nord04} used
\Hipp\ positions and proper motions, supplemented by spectroscopic
observations, to construct a homogeneous sample of nearby mostly F and
G dwarf stars.  We employ their updated catalog \citep{Holm09}
that uses their revised temperature calibration and the improved
astrometry from the reanalysis of \Hipp\ data by \cite{vL07}.  This
monumental effort has resulted in a large sample of stars in the solar
neighborhood with distances and full space motions.  \cite{Holm09} do
not supply individual uncertainties for each velocity, but assert that
they are believed accurate to 1.5~km~s$^{-1}$, with the greatest
contribution coming from distance uncertainties.

The machine readable table of $16\,682$ stars made available by
\cite{Holm09} includes sky positions, distances, and the $(U, V, W)$
components of the star's motion relative to the Sun in Galactic
coordinates.\footnote{These Cartesian velocity components are oriented
  such that $U$ is towards the Galactic centre, $V$ is in the
  direction of Galactic rotation, and $W$ is towards the north
  Galactic pole.}  We discard those having no distance, no $(U,V,W)$
velocities, or no estimate of \Teff.  We do not use the disputed age
estimates \citep[\eg][]{Reid07,Holm07} in the present paper.  We
correct for Solar motion relative to the local standard of rest (LSR)
by adding $(U,V,W)_\odot = (11.10,12.24,7.25)\;$km~s$^{-1}$
\citep{Scho10} to the tabulated velocities.

\citet{Nord04} note that they obtained \FeH$\;\ga0.4$ for a small
number of stars, though none is within 40~pc of the Sun.  Furthermore,
these high values are found only for stars with \Teff$\;\ga 6200\;$K.
They argue that the coincidence of high \Teff\ and \FeH, together
with other information, suggests that the extinction to these stars
has been overestimated.  We therefore discard a further 29 stars
having \FeH$\;> 0.4$.

In order to select nearby disk stars, we further restrict the sample
to stars whose best estimate of the distance is within $500\;$pc,
$|U|,\;|V|< 80\;$km~s$^{-1}$, and retain only those having an energy
of vertical motion about the Galactic mid-plane, $E_z = 0.5(z^2\nu^2 +
W^2) < 392\;$(km~s$^{-1})^2$, with the vertical frequency $\nu =
0.07\;$km~s$^{-1}$~pc$^{-1}$ \citep{BT08}, giving them a maximum
vertical excursion of $\pm 400\;$pc.  Thus we select only stars that
have a high probability of being thin disk stars, in the same spirit
as \cite{Bens03}, but not in exactly the same manner\footnote{Adopting
  their selection criterion for thin disk stars, changes the number of
  selected stars by a few hundred, but our conclusions are unaffected
  by this marginal revision.} (see \S\ref{discussion} for further
discussion of selection effects).  Our final sample contains $11\,877$
stars that we use in this analysis.

\begin{figure}[t]
\begin{center}
\includegraphics[width=.9\hsize,angle=0]{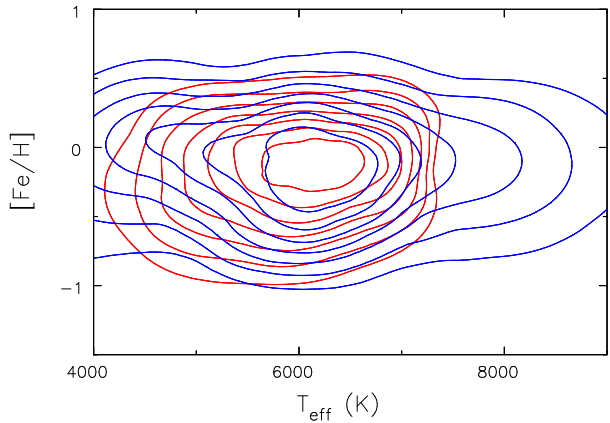}
\end{center}
\caption{The densities of stars in the space of \Teff\ and \FeH\ in
our selected \GCS\ (red) and \RAVE\ (blue) subsamples.}
\label{TevsFe}
\end{figure}

The median distance of the selected stars is 75~pc and the sample
within 40~pc of the Sun is believed to be near complete.  We adopt
$R_0=8\;$kpc, the circular speed of the LSR to be $V_0 =
220\;$km~s$^{-1}$, and compute the Galacto-centric angular momentum,
$L_z$, using the in-plane distance of the star from the Sun and the
peculiar velocities $(U,V)$ corrected for the Sun's motion.

The red contours in Fig.~\ref{TevsFe} show the distribution of the
selected \GCS\ stars in the space of \Teff\ and \FeH.  The range of
\Teff\ indicates that the sample extends outside the spectral types F
and G, reflecting the selection criteria described by \citet{Nord04}.
These authors estimate uncertainties of $\la 0.1$ in \FeH\ and 94~K
in \Teff.

\citet{Casa11} present a recalibration of \Teff\ and \FeH\ for the
\GCS\ stars that folds in the infrared flux from a star.  They obtain
slightly higher values for \Teff\ and \FeH\ than those derived by
\citet{Holm09} and they too attempt to assign ages for the individual
stars.  The additional information used by \citet{Casa11} should lead
to improved values for the derived quantities, but they claim reliable
results for fewer than half the \GCS\ stars (their {\it irfm\/}
sample) and for none of the cooler stars.  We therefore continue to
use the calibration for the whole \GCS\ sample by \citet{Holm09}.

\subsection{RAVE sample}
\label{RAVEsample}
The \RAVE\ survey \citep{Stei06} plans to measure the heliocentric
radial velocities and stellar parameters for about a million stars in
the southern sky having apparent magnitudes in the range $9 < m_I <
12$; the first $77\,461$ are available in the third data release
\citep{Sieb11}.  The typical uncertainty in the radial velocity is
$<2\;$km~s$^{-1}$, but the distance to most stars has to be judged
photometrically and most proper motions are from ground-based data.
Thus three of the phase space coordinates for each star are of much
lower quality than are those in the \GCS, although this weakness will
be compensated by a much larger sample size when the survey is
complete.

We have downloaded the on-line table of the third data release from
the \RAVE\ website and selected a subset of stars for
analysis.\footnote{The paper by \citet{Cosk11} presenting an analysis
  of the radial abundance gradient from these same stars appeared
  while this paper was in review.}  We estimate distances to these
stars by fitting to the Yonsei-Yale isochrones \citep{Dema04} using a
method related to that described by \citet[][see also Burnett
  \etal\ 2011 for an alternative method]{Bred10}.  We adopt many of
their selection criteria: we require the spectral signal-to-noise
parameter S2N$\; >20$ with a blank spectral warning flag field; the
parameters [M/H], log($g$), and \Teff\ to be determined; and the stars
to have J and K$_s$ magnitudes from 2MASS with no warning flags about
the identification of the star or the 2MASS photometry.  Unlike those
authors, however, we have kept stars with $b<25\degr$ on the grounds
that extinction for the nearby stars that interest us will not be
large enough in the near IR to severely bias our distance estimates.
As we are here interested only in nearby main-sequence stars that are
members of the disk population, we also make preliminary cuts to
eliminate stars with log$(g)<4$, $T_{\rm eff}> 10^4\;$K, and with
$|v_r|>120\;$km~s$^{-1}$.  We have treated the tabulated [M/H] as
equivalent to \FeH\ since the corrections \citep{Zwit08} are generally
within the uncertainties.  We also show in \S2.3 that the [M/H] values
for \RAVE\ stars agree with the \FeH\ values for the stars in common
with the \GCS.

We estimate the absolute J magnitude of each selected star by matching
the estimated \FeH, log($g$), \Teff, and \hbox{J-K$_s$} color to
values in the isochrone tables for stars of all ages and all values of
[$\alpha$/Fe], rejecting a few more stars for which the best match
$\chi^2>6$.  We consider the closest match in the tables to the given
input parameters to yield the best estimate of the absolute magnitude
from which we estimate a photometric distance using the apparent
J-band magnitude.\footnote{\citet{Zwit10} describe a similar method,
  but define a ``most likely'' estimate of the absolute magnitude that
  differs from our ``best'' estimate.  The difference is likely to be
  well within the uncertainties for the main-sequence stars considered
  here.}  We save the values of \FeH, log($g$), \Teff, and J-K$_s$
color of the closest matching model star in the isochrone table;
Monte-Carlo variation of the stellar parameters about this saved set
of values suggests that distances have a fractional precision of 30\%
-- 50\%, with some larger uncertainties.

We use the proper motions in equatorial coordinates tabulated in
\RAVE, mostly from Tycho-2 \citep{Hog00},\footnote{We have not used
  the newly available proper motions from the fourth release of the
  Southern Proper Motion Catalog \citep{Gira11}} which we then combine
with the radial velocity and position to determine the heliocentric
velocity in Galactic components $(U, V, W)$ \citep{JS87,Piatek}.  We
estimate uncertainties in these velocities from 500 Monte Carlo
re-selections of all the stellar parameters that affect the distance
estimate, adopting $\sigma({\rm J}) = 0.03\;$mag, $\sigma({\rm J-K}_s)
= 0.042\;$mag, $\sigma(T_{\rm eff})=300\;$K,
$\sigma(\log\,g)=0.3\;$dex, and $\sigma([{\rm Fe/H}])=0.25\;$dex
\citep{Bred10}, as well as the tabulated radial velocity and proper
motion uncertainties.

We exclude stars more distant than 500~pc, correct the space
velocities for solar motion and apply the same restrictions as for the
\GCS\ sample to select only those with a high probability of being thin
disk stars.

\begin{figure}[t]
\begin{center}
\includegraphics[width=\hsize,angle=0]{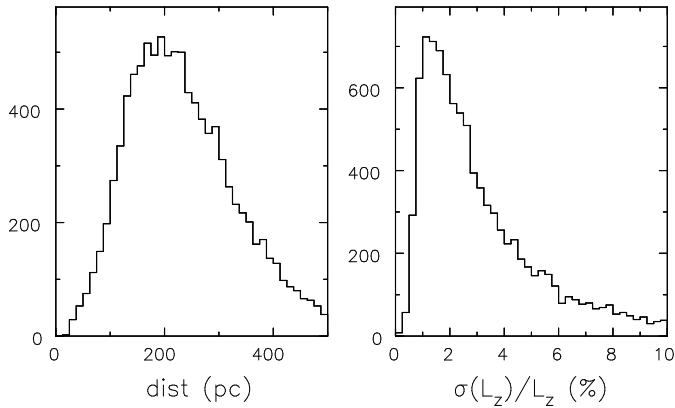}
\end{center}
\caption{Histograms of distances and $L_z$ uncertainties for the
  selected sample of 9803 \RAVE\ stars.}
\label{RAVEsel}
\end{figure}

Fig.~\ref{RAVEsel} shows histograms of distances and of uncertainties
in $L_z$ for the $9,803$ remaining stars.  The median distance is
220~pc from the Sun and uncertainties in $L_z$ are generally smaller
than 5\%.

The blue contours in Fig.~\ref{TevsFe} show the distribution of
selected \RAVE\ stars.  The distribution in \FeH\ values is slightly
broader than for the \GCS\ stars, reflecting in part at least the
lower precision of the spectra.  However, the rms scatter in
\FeH\ appears to be $\sim 0.2$, suggesting the uncertainty of 0.25
estimated by \citet{Sieb11} is somewhat pessimistic for this
sub-sample of main-sequence stars.  These authors also suggest
uncertainties of 300~K in their estimates of \Teff, and the broader
range of \Teff\ than in the \GCS\ sample is therefore real.

\subsection{Stars in common}
The overlap between the \GCS\ and \RAVE\ samples of stars is quite
small but, of those we have selected, precisely 100 stars are
positionally coincident.  This small degree of overlap allows us to
compare the spectral parameters, \Teff, \FeH, and radial velocity
($v_r$), derived separately by the two teams from their independent
spectra.

We find that differences in the estimated $v_r$ for the same stars
average just $0.32\;$km~s$^{-1}$ higher in \RAVE\ than in \GCS, with
an rms scatter about this mean of only $2.3\;$km~s$^{-1}$.  Similarly,
differences in \Teff\ have a mean of 203~K, again higher in
\RAVE\ than in \GCS, with an rms scatter about the mean of 202~K.
Finally, the mean difference in \FeH\ is 0.057, lower in \RAVE\ than
in \GCS, with an rms scatter of 0.18.  We have looked in vain for
systematic variations of these differences, which appear to be
consistent with random scatter.

Thus all three spectral parameters agree between the independent
measures to an accuracy that is significantly better than expected
from the quoted uncertainties, indicating that both teams have been
very careful with their calibrations and conservative with their
uncertainty estimates.

\begin{figure}[t]
\begin{center}
\includegraphics[width=\hsize,angle=0]{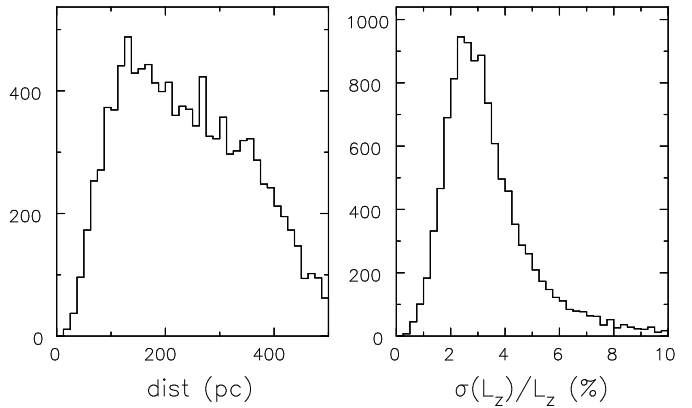}
\end{center}
\caption{Histograms of distances and $L_z$ uncertainties for the
  selected sample of $11\,022$ M-dwarf stars.}
\label{Mdsel}
\end{figure}

\subsection{M-dwarf sample}
\label{Mdsample}
The \SDSS\ \citep{York00} and Segue2 \citep{Yann09} surveys are
complete, but sample fainter stars than \RAVE\ (the magnitude range
for Segue2 was $14.0 < g < 20.3$) that are therefore generally more
distant.  Since the Sloan spectral parameters pipeline \citep{Pipe}
does not attempt to fit stars with $T_{\rm eff} \la 4500\;$K, almost
all main sequence stars with estimated parameters in \SDSS\ DR8 are
more distant than 500~pc.

Fortunately, \cite{West11} provided a catalog of 70\,841 M-dwarfs from
\SDSS\ DR7.  Their table provides a photometric estimate of the
distance to each star, as well as a spectroscopic radial velocity, and
proper motion from USNO-B/SDSS catalog \citep{Munn04,Munn08}.  They
suggest that distance uncertainties are typically about 20\% and
uncertainties in radial velocity are 7-10~km s$^{-1}$.  None of these
stars are in \RAVE\ (because they are in different parts of the sky)
and all are much cooler than the \GCS\ stars.

As \cite{West11} recommend, we selected stars with the ``goodPM'' and
``goodPhot'' flags set to `true', and the ``WDM'' flag set to `false'
to eliminate possible binaries with a white dwarf companion, which
reduces the sample to 39\,151 stars.  We further excluded stars having
no radial velocity as well as those with no distance estimate or for
which the estimated distance exceeded 500~pc.  As for the \GCS\ and
\RAVE\ stars, we correct the space velocities for solar motion and
apply the same restrictions to select only those with a high
probability of being thin disk stars, leaving us with a final sample of
$11\,022$ stars.

We estimated uncertainties in the velocity components, $(U,V,W)$ by
combining the 20\% distance uncertainty, a 10~km s$^{-1}$ uncertainty
in the radial velocity, and the proper motion uncertainties.  The
distribution of distances and fractional $L_z$ uncertainties is shown
in Fig.~\ref{Mdsel}; many of these intrinsically faint stars lie
within 200~pc and, again, angular momentum uncertainties are typically
$\la 5$\%.

As estimates of \Teff\ or \FeH\ are not easily derived from low
resolution spectra of such cool stars, \cite{West11} do not attempt to
provide these quantities in their catalog.  Instead, they provide
spectral sub-class, which can be related to \Teff, and a titanium
oxide index, $\zeta_{\rm TiO}$, which is believed to be an indicator
of metallicity.

\section{Results}
\subsection{GCS and RAVE stars}
\begin{figure}[t]
\begin{center}
\includegraphics[width=\hsize,angle=270]{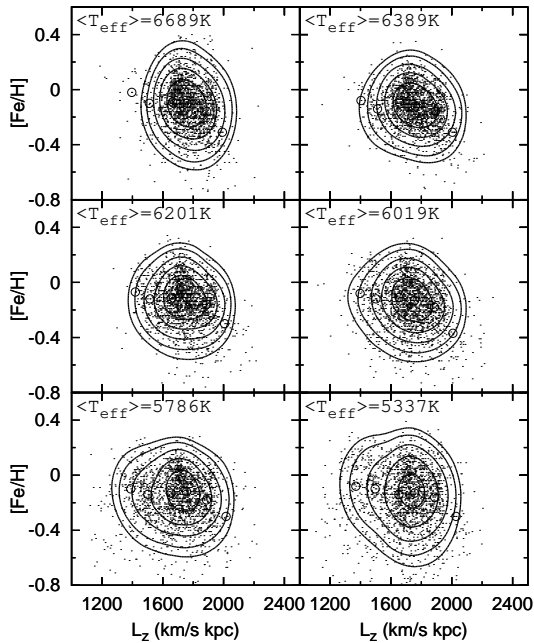}
\end{center}
\caption{The distributions of \GCS\ stars in the space of $L_z$ and
  \FeH\ when grouped by effective temperature.  The temperature bins
  were chosen to contain equal numbers of stars and the average
  temperature of the stars in each bin is given.  The large open
  circles, which are purely to illustrate the trends, show the median
  \FeH\ for stars with $1300 < L_z < 2100$ divided into six bins in
  each panel.  The unit of angular momentum is km~s$^{-1}$~kpc.}
\label{GCScont}
\end{figure}

\begin{figure}[t]
\begin{center}
\includegraphics[width=\hsize,angle=270]{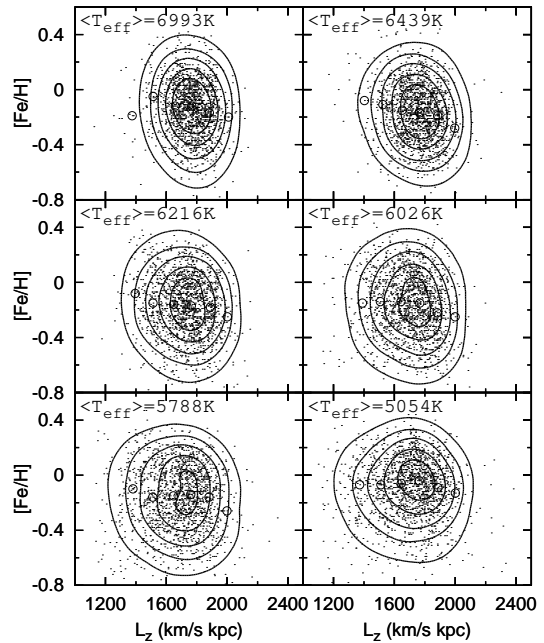}
\end{center}
\caption{Same as Fig.~\ref{GCScont}, but for the \RAVE\ sample.}
\label{RAVEcont}
\end{figure}

Figs.~\ref{GCScont} \& \ref{RAVEcont} show the metallicity
\vs\ angular momentum of \GCS\ and \RAVE\ stars each separately binned
into various temperature ranges.  The bins are chosen so as to have
equal numbers of stars, and the average \Teff\ of the stars in each
panel is shown.  The density of points in each panel is indicated by
the contours.

If the rotation curve of the Milky Way were flat at a constant
circular speed of 220~km~s$^{-1}$, then stars with $L_z =
1\,210\;$km~s$^{-1}$~kpc would have home radii of 5.5~kpc, while the
home radii of those with $L_z =2\,200\;$km~s$^{-1}$~kpc would be
10~kpc.  (Note that these numerical values will simply scale with our
choices of $R_0$ and $V_0$.)  Thus, although the stars in our sample
are passing close to the Sun right now, many are visiting from quite
far afield.


A number of trends can be seen: The spread in $L_z$ is greater for the
cooler stars, reflecting the fact that the higher velocity dispersions
of older stars allows stars having a wider range of $L_z$ to visit the
solar neighborhood.  The rms scatter in \FeH\ is in the range $0.19
\la \sigma \la 0.23$ for the \RAVE\ stars in each panel, which is less
than the nominal uncertainty, as noted in \S\ref{RAVEsample}.  The
spread in \FeH\ for the \GCS\ stars is generally smaller, rising from
0.14 in the second panel to 0.21 for the coolest stars.  The spread
for the hottest \GCS\ stars (0.16) is anomalously high, due to
incomplete removal of stars that were over-corrected for reddening, as
noted in \S\ref{GCSsample}.  While the large spread in \FeH\ is partly
due to uncertainties and partly real, each panel has a weak declining
trend of metallicity with increasing $L_z$ that reflects the
metallicity gradient in the Galaxy.  We have added the open circles,
which show the variation of the median in bins of $L_z$ in each panel,
simply to make this last trend more clearly visible.

\subsection{Trend in gradient}
\begin{figure}[t]
\begin{center}
\includegraphics[height=\hsize,angle=270]{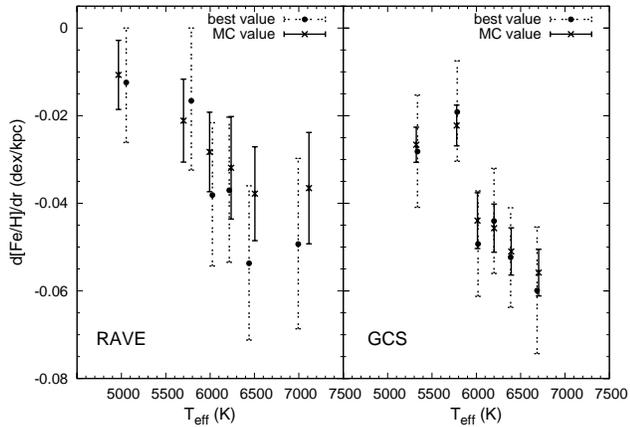}
\end{center}
\caption{The metallicity gradient, $d$\FeH$/dL_z$ estimated from the
  \GCS\ sample shown in Fig.~\ref{GCScont} (right panel), and from the
  \RAVE\ sample in Fig.~\ref{RAVEcont} (left panel).  The filled
  symbols show the best estimate from Theil's algorithm, with 95\%
  confidence limits (dashed), the crosses with solid error bars show
  the standard errors derived from a Monte Carlo simulation with the
  adopted uncertainties.}
\label{slp}
\end{figure}

We adopt Theil's non-parametric method \citep{HW99} to quantify the
metallicity gradient in each panel, which we plot as filled symbols in
Fig.~\ref{slp}, with the dashed error bars showing the 95\% confidence
limits.  This method returns the median slope between every pair of
points in the sample with the confidence limits being determined by the
range of the pairwise values.  We have converted the gradient with
angular momentum to the more usual dex per kpc by multiplying by our
adopted circular speed $V_0 = 220\;$km~s$^{-1}$.  Our abundance
gradient estimates therefore assume only a locally flat rotation curve
and a distance scale set by adopting $R_0=8~$kpc.

While these estimates of the metallicity gradient are derived from
stars that happen to be close to the Sun at the present time, the
sample includes stars from a broad range of home radii, as noted
above.  Stars whose ages greatly exceed the radial epicycle period,
$\sim 170\;$Myr at the solar radius, should be uniformly distributed
in radial phase around their epicycles.  Thus both the spread in
metallicities of the stars in our sample and the gradient we measure
are characteristic of the stars at their home radii, and our estimates
of the metallicity gradient are based on a radial range of several
kiloparsec.

The apparent flattening of the metallicity gradient with decreasing
\Teff\ is suggested by the \RAVE\ stars (left panel of Fig.~\ref{slp})
and is more convincing in the better-quality data of the \GCS\ sample
(right panel).  As this is our main result, we use Monte-Carlo
simulations to check whether the uncertainties in the slopes are small
enough to justify our claim of a trend.  We generate 100 new samples,
by randomly resampling stars, adding normally distributed values to
\Teff, \FeH, and $L_z$ to simulate measurement errors.  The size of
the errors are determined by the appropriate uncertainties from each
survey as described above or, for the $L_z$ values of the
\RAVE\ stars, by propagating our own estimates of the distance, proper
motion and radial velocity uncertainties.  We divide each sample into
equal numbers of stars in successive bins of \Teff\ and compute the
slopes.  The means and standard errors of these re-estimates are also
shown by the crosses and solid error bars in Fig.~\ref{slp}.  Note
that the solid error bars from our Monte-Carlo simulations indicate
the 1-$\sigma$ dispersion, whereas the dashed error bars from the
Theil algorithm show the 95\% confidence interval; the two independent
estimates are therefore consistent, and the flattening of the slope
with decreasing \Teff\ seems to be confirmed.  Note also that the
somewhat shallower trend in the Monte-Carlo re-estimates from the
\RAVE\ data is expected, since resampling the data in this manner
inserts additional uncertainties over and above those already present
in the data, which inevitably weakens a trend.  This effect is more
marked in the \RAVE\ sample because uncertainties in their data are
larger.

It also occurred to us that the change of slope could perhaps be a
systematic effect due to the above-noted increasing spread in $L_z$
with decreasing \Teff\ -- \ie, the greater spread in $L_z$ could
simply make the fitted slope appear shallower, while in fact the data
may be consistent with a constant slope.  We tested for this
possibility by creating some pseudo star samples that adopt a linear
relation between \FeH\ and $L_z$ that is the same for all the stars in
either the \RAVE\ or the \GCS\ sample, irrespective of their
temperature, which is the null hypothesis that we argue Fig.~\ref{slp}
disproves.  For each star in the sample, we use the measured $L_z$ to
assign a new \FeH\ from a Gaussian distribution about the mean from
the adopted trend with a dispersion equal to that about the linear
regression line for the entire sample.  We then bin this pseudo-sample
by \Teff\ as before and test whether the cooler stars, that have the
larger spread in $L_z$, have a shallower apparent slope.  With 500
realizations for each of the \GCS\ and \RAVE\ samples, we confirmed
that we recover the original input slope in every bin.  The absence of
any systematic trend with \Teff\ in this test increases our confidence
that the trend we find in the data in Fig.~\ref{slp} is real.

\begin{figure}[t]
\begin{center}
\includegraphics[height=\hsize,angle=270]{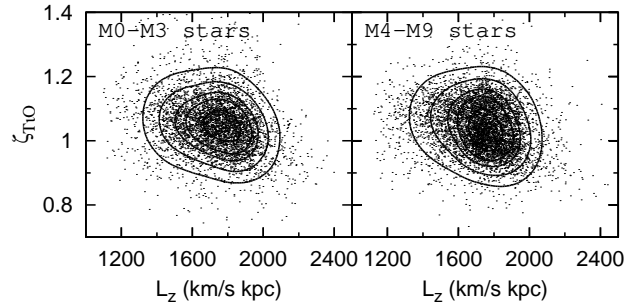}
\end{center}
\caption{TiO index, $\zeta_{\rm TiO}$ \vs\ angular momentum for the
  M-dwarf sample, divided by spectral sub-type into two roughly
  equally numerous groups.}
\label{Mdwcont}
\end{figure}

\begin{figure}[t]
\begin{center}
\includegraphics[width=\hsize,angle=270]{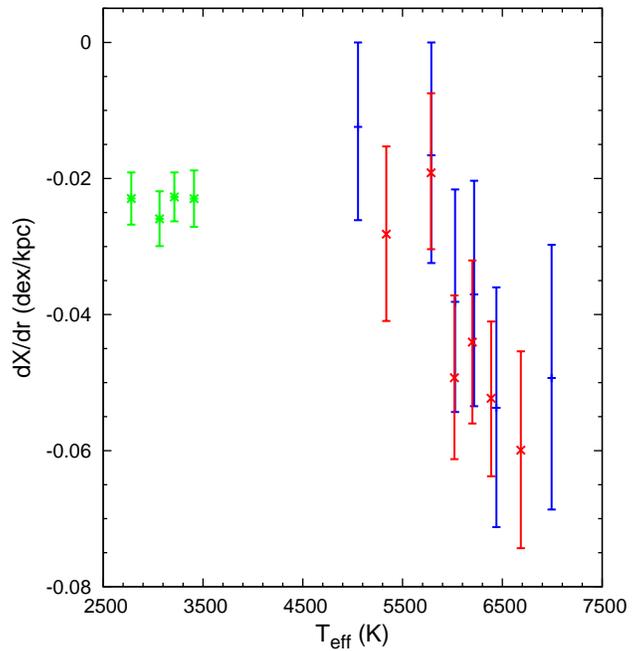}
\end{center}
\caption{The estimated metallicity gradient $d$\FeH$/dL_z$ from the
  \GCS\ stars (red) and from the \RAVE\ stars (blue) reproduced from
  Fig.~\ref{slp}.  The green points show $d\zeta_{\rm TiO}/dL_z$ from
  the M-dwarf stars in Fig.~\ref{Mdwcont} further subdivided by
  spectral sub-class.  The error bars show 95\% confidence limits
  from Theil's algorithm.}
\label{combine}
\end{figure}

Furthermore, both samples also hint at a break in the trend with
temperature around \Teff$\; \la 6\,000\;$K, although these points
alone are also consistent with a uniform trend.  In order to find
extra evidence for this possible break of slope, we attempt to include
the M-dwarf stars into our analysis.

\subsection{M-dwarf sample}
Fig.~\ref{Mdwcont} shows the distribution of $L_z$ and $\zeta_{\rm
  TiO}$ for the M-dwarf stars separated into two groups by spectral
class.  This metallicity indicator again has a noticeable gradient
with $L_z$.

As for the hotter stars, we have used Theil's algorithm to estimate
the gradient $d\zeta_{\rm TiO}/dL_z$, dividing the M-dwarf sample into
four groups by stellar sub-class, and plot the results in green in
Fig.~\ref{combine}.  (The abscissae were determined from the relation
\Teff$\; = 3700 - 150S\;$K, where $S$ is the spectral sub-class,
fitted by eye to the compilation by \citet{Reyl11}.)  The M-dwarfs
have a constant metallicity gradient that appears independent of
\Teff.

Because they are measured from a different quantity, the values of
$d\zeta_{\rm TiO}/dL_z$ cannot be compared directly with those of
$d$\FeH$/dL_z$ from the hotter stars, shown in red and blue.  We tried
using the empirical relation between $\zeta_{\rm TiO}$ and
\FeH\ suggested by \citet{Wool09}, but many \FeH\ values required an
extrapolation beyond the range calibrated by their study, and the
resulting mean \FeH\ is unreasonably high.  Thus we lack a direct
basis to relate the green points in Fig.~\ref{combine} to the others.

However, we note that if $\zeta_{\rm TiO}$ is proportional to \FeH,
the gradient $d$\FeH$/dL_z$ will scale with $d\zeta_{\rm TiO}/dL_z$,
and we would merely need to shift the green points vertically by some
fixed factor to bring them onto the same scale.  Since an
anti-correlation between $\zeta_{\rm TiO}$ and \FeH\ seems unlikely,
the green points probably could not be shifted above $d$\FeH$/dL_z =
0$.  Thus the data from the M-dwarf stars do seem to confirm that the
decreasing trend of metallicity gradient with \Teff\ among the hotter
stars really does break to no trend around \Teff$\; \la 6\,000\;$K,
\ie\ for stars whose main-sequence lifetimes are $\sim 10^{10}\;$yr.

\section{Discussion}
\label{discussion}
Many authors \citep[\eg][]{Lee11,Casa11,Cosk11} express concerns that
estimates of the metallicity gradient from a sample of stars depend
upon the adopted selection criteria.  This is because metallicity is a
function of age and, relative to younger stars, older stars typically
have larger peculiar velocities, which causes an asymmetric drift and
greater vertical motion.  This well-known kinematic trend must cause
an age bias in kinematically selected samples.

\subsection{Possible selection effects}
In this work, we apply kinematic and vertical motion cuts to eliminate
most thick disk and halo stars, which may well also eliminate some old
thin disk stars that should perhaps be included in our samples.
However, any bias in our analysis created by excluding older stars
because of their higher peculiar velocities will cause lower
temperature bins, which are the only bins to include stars of the old
disk, to have unrepresentatively few older disk stars.

Radial mixing implies that these possibly-missing stars would be
members of the most well-mixed population with the lowest radial
metallicity gradient, and adding significant numbers of them to the
lower \Teff\ bins would only increase the trend we observe.  In fact,
we find that the trend of metallicity gradient with \Teff\ is little
affected by limiting the vertical excursions of selected stars to
either 300~pc or 500~pc, suggesting that this cut is not biasing our
result.  Further restricting the vertical oscillations to $< 300\;$pc,
on the other hand, does exclude a disproportionate fraction of older
stars and the trend with \Teff\ is weakened, as our argument predicts.
Naturally also, the number of selected stars decreases with a more
severe restriction increasing the statistical uncertainty in our
estimate of the metallicity gradient.

We do not use any measure of chemical composition to select stars, and
our sample may therefore include some stars with enhanced
[$\alpha$/Fe].  Whether such abundances best characterize thick disk
stars is the subject of current debate \citep{Ivez08,Lee11} while the
existence of a thick disk that is distinct from the thin has been
questioned \citep{Bovy11}.  Our study is unaffected by these
questions, because we are interested only in how the metallicity
gradient changes with the mean age of the stars and, provided we have
excluded halo stars that are little affected by spiral activity, we do
not need to consider whether each star may or may not belong to a
particular population.  Note that \citet{SSS12} have shown that mixing
is only gradually weakened by increased random motion and vertical
thickness.

A separate issue is that all the stars in this study are close to the
Sun at the present time, and we therefore cannot include stars with
home radii far from the Sun but with radial excursions that are too
small to bring them into the solar neighborhood.  However, provided
the stars we do see have abundances that are representative of those
of their \Teff\ and home radius, the omitted stars will not affect our
estimate of the slope.  Since we are not selecting on metallicity,
there should be no bias in our slope estimates.

\subsection{Comparison with other work}
\citet{Casa11} re-analysed the \GCS\ using a revised calibration of
\Teff\ and \FeH\ and new estimates of the individual stellar ages.
We have tried to compare our results with theirs, which is not
straightforward because our bins of \Teff\ each contain stars having a
broad range of ages.  We have used isochrone tables \citep{Dema04} to
estimate the main-sequence lifetimes of the individual stars in the
\GCS\ sample, using the \Teff\ and \FeH\ of each, and so find the
average main-sequence lifetime of the stars in our separate
\Teff\ bins.  The mean maximum age of stars defined in this way
increases from about 2 Gyr for our hottest bin to 10 Gyr for the
coolest.  Using these mean maximum ages to replot the trend of
metallicity gradient in our Fig.~8, we find reasonable agreement with
the gradient values as a cumulative function of age shown in
Fig.~18(d) of \citet{Casa11}.

Our estimated gradient is in the range of other measurements from
stars.  For example, \cite{Cosk11} derived a metallicity gradient of
-0.051 dex kpc$^{-1}$ for F-dwarfs, and they also find a shallower
gradient for the slightly older G-dwarfs.  While their sample clearly
overlaps with ours, they adopted different selection criteria and
employed a different technique to determine the home radii, it is
encouraging that they reached similar conclusions.  Using disk
Cepheids, which are young objects, \citet{Luck11} derived \FeH$ =
-0.055R + 0.475$, where $R$ is the Galacto-centric radius, again in
tolerable agreement with our gradient estimate for the hotter stars.

\citet{Chen03} estimate a radial gradient of \FeH\ of $-0.063\pm0.008$
dex kpc$^{-1}$ from 118 star clusters.  Both they and \citet{Frie02}
report that, when the clusters are divided by age, the slope appears
steeper for the older sample, which is the opposite of our finding
here.  Note that the radial distribution of star clusters in these
papers extends outwards from the solar radius, whereas our sample of
stars overlaps the solar radius and extends inwards to $\sim 5\;$kpc
-- thus the different variations with age could indicate the absence
of mixing in the outer disk.  However, other systematic differences
may complicate this interpretation; \eg, the outer disk open clusters
in the sample assembled by \citet{Chen03} are farther from the
mid-plane than are those near the solar radius.

\citet{Maci06} also claim a steeper abundance gradient for older
planetary nebulae, although their estimated gradients for the
different elements show a lot of scatter and ages are less reliable
than for star clusters.  It should be noted that \citet{SH10} reached
the opposite conclusion finding, for a sample of PNe concentrated over
the radial range $3< R < 12\;$kpc, a shallower abundance gradient for
the older PNe, which is in better agreement with our results.

\section{Conclusions}
We have used samples of local thin-disk, main-sequence stars grouped
by effective temperature to demonstrate that the radial abundance
gradient in the disk of the Milky Way flattens as the \Teff\ of the
stars in the sample decreases.  We eliminate the blurring effect of
epicyclic motion by estimating the angular momentum, $L_z$, of each
star about the center of the Galaxy, since the home, or guiding
center, radius of a star is determined only by its angular momentum.
Although the stars in our sample are all within 500 pc of the Sun,
their home radii are spread over a wide swath of the Milky Way disk.

We interpret \Teff\ as a proxy for mean age, and conclude that a
shallower gradient for stars of a greater mean age.  We find some
evidence that the trend of decreasing metallicity gradient with
increasing age breaks to a flat trend around $5\,500 \la \hbox{\Teff}
\la 6\,000\;$K, for which the main-sequence lifetime of a star is
approximately 10~Gyr, or roughly the age of the thin disk.  Flattening
of the metallicity gradient is consistent with the predictions of
radial mixing, or churning, due to spiral patterns in the disk.

\acknowledgments We thank Birgitta Nordstr\"om for helpful
correspondence about the apparently metal rich stars in the \GCS\ and
an anonymous referee for comments that have helped us to improve the
paper.  This work was supported by NSF grant AST 09-37523 (PI
Newberg), and grant 11061120454 (PI Deng) from the National Science
Foundation of China (NSFC), the NSFC grants 11073038 (PI Chen),
11173044 (PI Hou), Key Project 10833005 (PI Hou), and by the Group
Innovation Project No. 10821302.


\end{document}